\documentclass[12pt]{article}

\usepackage{graphics}

\title{Mean-atom-trajectory model for the velocity autocorrelation function 
       of monatomic liquids}
\author{Eric D.\ Chisolm, Brad E.\ Clements, and Duane C.\ Wallace \\ 
        Theoretical Division \\ Los Alamos National Laboratory \\ Los Alamos, 
        NM~~87545}

\begin{document}

\maketitle

\begin{abstract}
We present a model for the motion of an average atom in a liquid or
supercooled liquid state and apply it to calculations of the velocity
autocorrelation function $Z(t)$ and diffusion coefficient $D$.  The
model trajectory consists of oscillations at a distribution of
frequencies characteristic of the normal modes of a single potential
valley, interspersed with position- and velocity-conserving transits
to similar adjacent valleys.  The resulting predictions for $Z(t)$ and
$D$ agree remarkably well with MD simulations of Na at up to almost three
times its melting temperature.  Two independent processes in the model
relax velocity autocorrelations: (a) dephasing due to the presence of
many frequency components, which operates at all temperatures but
which produces no diffusion, and (b) the transit process, which
increases with increasing temperature and which produces diffusion.
Because the model provides a single-atom trajectory in real space and
time, including transits, it may be used to calculate all single-atom 
correlation functions.
\end{abstract}

\section{Introduction}
\label{intro}

In order to explain the experimental fact that the specific heat of a
solid changes little, while its self-diffusion coefficient changes
greatly, when it melts to a liquid, Frenkel \cite{frenk1,frenk2}
suggested that an atom in a liquid undergoes approximately harmonic
vibrations about an equilibrium position, occasionally jumping from
one equilibrium position to another; these jumps are responsible for
self-diffusion.  Using molecular dynamics calculations, Stillinger and
Weber (\cite{stillweb1}-\cite{stillweb4}) demonstrated the presence of
local many-particle minima (``inherent structures'') in the potential
surface underlying the liquid state, and they observed that
``diffusion and fluid flow within a liquid may be interpreted as
transitions between $\ldots$ local minima'' \cite{stillweb3}.
Building on these ideas, Zwanzig \cite{zwan} studied the
self-diffusion coefficient $D$, given in terms of the velocity
autocorrelation function $Z(t) = \frac{1}{3} \langle \mbox{\boldmath
$v$}(t) \cdot \mbox{\boldmath $v$}(0) \rangle$ by the Green-Kubo
formula \cite{hansmcd}
\begin{equation}
D = \int_{0}^{\infty} Z(t)\, dt.
\end{equation}
For harmonic motion about a many-particle equilibrium position, $Z(t)$ 
is given by 
\begin{equation}
Z(t) = \frac{kT}{M} \int \rho(\omega) \cos(\omega t)\, d\omega
\label{nondifZ}
\end{equation}
where $\rho(\omega)$ is the density of normal mode frequencies.  (The
derivation of this formula is discussed in \cite{zwan} and performed
in Section II of \cite{stratt6}.)  Zwanzig suggested that jumps
between equilibrium positions will have the effect of multiplying this
expression by a factor $\exp(-t/\tau_{\rm zw})$, where the ``hopping
time'' $\tau_{\rm zw}$ is characteristic of the time between jumps.
Much effort has been devoted to developing these ideas into full
harmonic theories of liquid dynamics, particularly theories of
self-diffusion in liquids and supercooled liquids
(\cite{1qnm}-\cite{buch}).  In most theories one finds $\rho(\omega)$
by expanding the potential energy to second order around each of some
set of configurations, diagonalizing the second-order term in the
potential (often called the dynamical matrix) to find the frequency
distribution for that configuration, and averaging over all
configurations chosen.  This is done in one of two different ways.  In
quenched normal mode (QNM) theories
\cite{1qnm,ohmine1,ohmine2,vijanit,rabgezber2, rabgezber4},
$\rho(\omega)$ is calculated at several potential minima and averaged,
while in instantaneous normal mode (INM) theories
(\cite{keyes1}-\cite{stratt12}, \cite{benlaird1}-\cite{wutsay4},
\cite{scior}, \cite{buch}) $\rho(\omega)$ is averaged over a thermal
distribution of configurations, with no special emphasis placed upon
configurations in potential valleys.  This difference manifests itself
in the fact that in INM theories, the configuration-averaged
$\rho(\omega)$ usually includes both real and imaginary frequencies
(corresponding to stable and unstable normal modes), since most
configurations will not lie at the bottoms of valleys, while in QNM
theories only real frequencies are represented.  In addition, the QNM
density of states for a given system will in general be
temperature-independent, while the INM density will depend strongly on
temperature.  The two most prominent ways to determine $\tau_{\rm zw}$
are (a) to extract it from the imaginary frequency INM distribution,
developed most notably by Keyes (\cite{keyes1}-\cite{keyes16}), and (b) 
to set $\tau_{\rm zw}^{-1}$ equal to the long-time decay rate of the
``cage correlation function'' of Rabani, Gezelter, and Berne
\cite{rabgezber2,rabgezber4}.  (Another theory uses Cao and Voth's
frequency-dependent multiplicative factor \cite{cv1,cv2}.)  Notice
that none of these theories attempt to model the actual motion of a
diffusing particle in the liquid and then calculate $Z(t)$ from this
motion; they try to model the effects of diffusion on the
autocorrelation function directly.

The theory of $Z(t)$ we propose differs from those discussed above
both in how we determine $\rho(\omega)$ and how we model the effects
of diffusion.  In Section \ref{model}, we suggest a density of states
of the QNM type that also incorporates insights drawn from studies of
the potential energy surface of liquid Na by Clements and Wallace
\cite{wall4,wall5}.  These studies were undertaken to test a theory of
monatomic liquid dynamics proposed by Wallace \cite{wall1} that has
been applied previously with some success to the thermodynamics of a
wide variety of liquid metals \cite{wall2} and an earlier study of
$Z(t)$ \cite{wall3}, and the present work is also intended to lend
credence to that theory.  We propose a model for the process of
diffusion based on the following observations.  The system moves in a
set of nearly harmonic many-particle valleys, and the motion within
each valley may be analyzed into normal modes of vibration about the
valley minimum.  The motion of the system from one valley to another
is called a {\em transit}, and it corresponds to a change in the
equilibrium positions of a small group of atoms.  When a transit
occurs involving a given atom, or one of its neighbors, the normal
mode eigenvector components for that atom will change, and since this
can happen many times during a single vibrational period (as we will
see in Section \ref{MD}), the eigenvectors will not necessarily
provide a useful basis for describing the motion.  This suggests that
an independent atom model will be a good theoretical starting point
\cite{wall3}. (Notice that this argument does not apply to INM, where
the eigenvectors change continuously with the motion, instead of
discontinuously only at transits.)  Therefore, we propose a
mean-atom-trajectory model which describes a single average particle
in the diffusing liquid actually transiting between single-particle
equilibrium positions, and from this model we calculate $Z(t)$
directly.  In Section \ref{MD} we compare our predictions with
molecular dynamics (MD) simulations of liquid Na over a very broad
range of temperatures, and in Section \ref{concl} we discuss our
results.

\section{The model}
\label{model}

\subsection{Density of states}
\label{DOS}

\begin{figure}[p]
\includegraphics{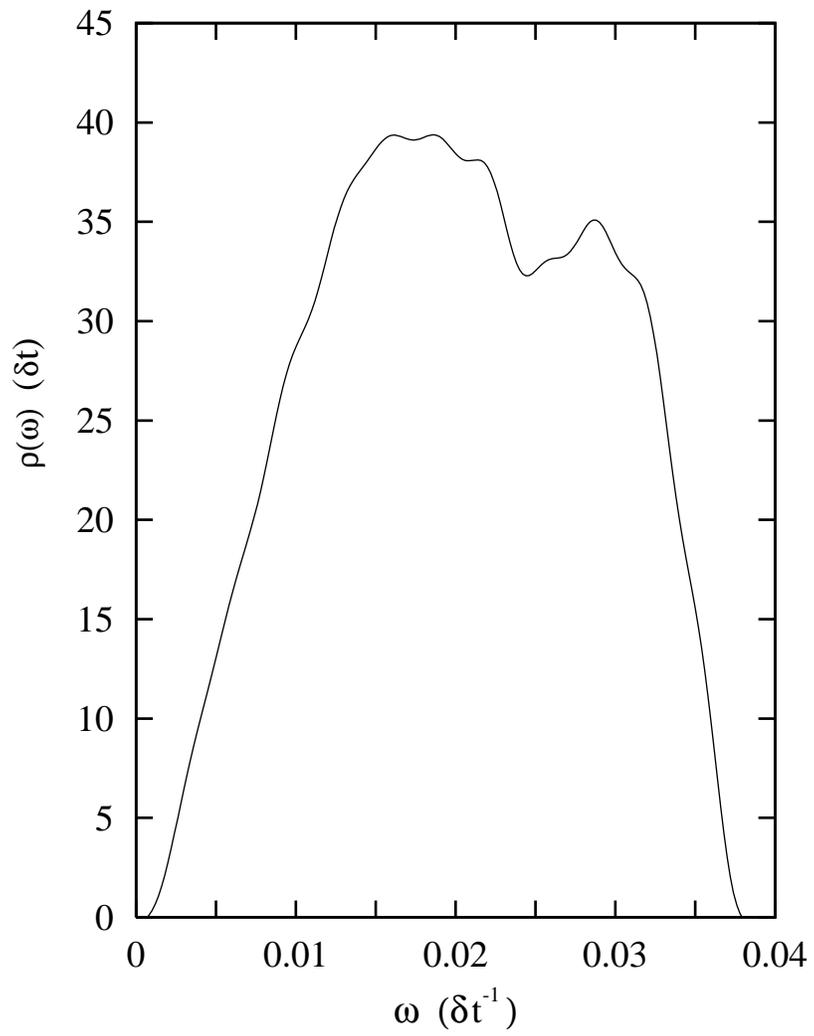}
\caption{$\rho(\omega)$ for liquid Na constructed from the set of 
         frequencies in Fig.\ 7 of \cite{wall4}.  Here $\delta t$
         is the MD timestep ($1.4 \times 10^{-15}$ s).}
\label{rhovsfreq}
\end{figure}

Wallace \cite{wall1} has predicted that in any monatomic liquid the
many-body potential valleys can be divided into three categories: the
few crystalline valleys; so-called ``symmetric'' valleys which retain
some remnant of the crystal system's symmetry; and ``random'' valleys,
in which no crystal symmetries remain.  He argued further that the
random valleys should greatly outnumber the symmetric ones (thus
controlling the statistical mechanics of the liquid), and that all
random valleys should have the same distribution of normal \mbox{mode}
frequencies.  Wallace and Clements \cite{wall4,wall5} have verified
these predictions for liquid Na, as well as discovering criteria that
one can use to determine whether the system is in a symmetric or
random valley when nondiffusing.  From Fig.\ 7 of \cite{wall4} one can
construct a distribution $\rho(\omega)$ for liquid Na that will be
valid whenever the system remains in a random valley; this
$\rho(\omega)$ is shown in Fig.\ \ref{rhovsfreq}.
However, since we actually have the set of normal mode frequencies
$\{\omega_{\lambda}\}$ found in \cite{wall4}, we can use them in Eq.\
(\ref{nondifZ}) directly, so our nondiffusing $Z(t)$ is
\begin{equation}
Z(t)  = \frac{1}{3N-3} \frac{kT}{M} \sum_{\lambda} \cos(\omega_{\lambda}t),
\label{realnondifZ}
\end{equation}
where $N$ is the number of particles in the system, and the number of
normal modes is $3N-3$ because the three zero-frequency modes
corresponding to center of mass motion are not excited.  This is our model 
$Z(t)$ when the system remains in a single random valley without diffusion.
Note that because all of the random valleys have the same $\rho(\omega)$, 
an average over quenched configurations is unnecessary.

\subsection{Motion of an Average Particle}
\label{MAT}

Our first goal is to construct a model for the motion of an average
particle that reproduces Eq.\ (\ref{realnondifZ}) for $Z(t)$ in the
absence of transits.  Since the system is transiting with overwhelming
likelihood from random valley to random valley, and since all the
random valleys have the same frequency distribution, it is sensible to
model an average particle's motion in terms of oscillations at those
frequencies, or
\begin{eqnarray}
\mbox{\boldmath $r$}(t) & = & \mbox{\boldmath $R$} + \mbox{\boldmath $u$}(t)
                              \nonumber \\
& = & \mbox{\boldmath $R$} + \sum_{\lambda} \mbox{\boldmath $w$}_{\lambda}
                             \sin(\omega_{\lambda}t + \alpha_{\lambda}),  
\label{rt}
\end{eqnarray}
where the particle's position $\mbox{\boldmath $r$}(t)$ is divided
into a center of oscillation $\mbox{\boldmath $R$}$ and oscillations
$\mbox{\boldmath $u$}(t)$ about that center, and the parameters in
$\mbox{\boldmath $u$}(t)$ aside from the $\omega_{\lambda}$ have yet
to be determined.  Let us assume that the values of the phases
$\alpha_{\lambda}$ are randomly distributed among the particles; then
one calculates $Z(t)$ from Eq.\ (\ref{rt}) by differentiating to find
$\mbox{\boldmath $v$}(t)$, computing the product $\mbox{\boldmath
$v$}(t) \cdot \mbox{\boldmath $v$}(0)$, and averaging over each of the
$\alpha_{\lambda}$ separately; the result is
\begin{equation}
Z(t) = \frac{1}{6} \sum_{\lambda} |\mbox{\boldmath $w$}_{\lambda}|^2 
       \omega_{\lambda}^2 \cos(\omega_{\lambda}t).
\label{kindaZ}
\end{equation}
Eq.\ (\ref{kindaZ}) becomes Eq.\ (\ref{realnondifZ}) with the choice
\begin{equation}
\mbox{\boldmath $w$}_{\lambda} = \sqrt{\frac{1}{N-1}\frac{2kT}
                                 {M\omega_{\lambda}^2}}
                                 \,\hat{\mbox{\boldmath $w$}}_{\lambda}
\label{wchoice}
\end{equation}
where $\hat{\mbox{\boldmath $w$}}_{\lambda}$ is an arbitrarily chosen
unit vector.  Thus Eq.\ (\ref{rt}) with the phases $\alpha_{\lambda}$
randomly chosen and $\mbox{\boldmath $w$}_{\lambda}$ given by Eq.\
(\ref{wchoice}), with the unit vectors $\hat{\mbox{\boldmath
$w$}}_{\lambda}$ also randomly chosen, constitute our mean-atom-trajectory 
model when the system is not diffusing.

To include diffusion in our model, we must incorporate both the rate
at which transits occur and their effect on the particle's motion.  We
will allow the rate $\nu$ to be a temperature-dependent parameter that
we will determine in Section \ref{MD} by fitting to MD simulations (so
the probability of a transit in small time $\Delta t$ is $\nu \Delta
t$), and we assume that the transit occurs instantaneously (the
particle simply crosses the surface separating distinct valleys), so
it must conserve both the particle's position $\mbox{\boldmath
$r$}(t)$ and velocity $\mbox{\boldmath $v$}(t)$.  To be more specific,
we assume that the transit occurs in the forward direction, so that
the center of the new valley lies an equal distance away from the particle 
but on the opposite side from the center of the old valley.  Let
$\mbox{\boldmath $r$}^{\rm before}(t)$, $\mbox{\boldmath $R$}^{\rm
before}$, and $\mbox{\boldmath $u$}^{\rm before}(t)$ be the position
parameters from Eq.\ (\ref{rt}) before the transit, and let
$\mbox{\boldmath $r$}^{\rm after}(t)$, $\mbox{\boldmath $R$}^{\rm
after}$, and $\mbox{\boldmath $u$}^{\rm after}(t)$ be the parameters
after; then our assumption of a forward transit implies that
$\mbox{\boldmath $u$}^{\rm after}(t) = -\mbox{\boldmath $u$}^{\rm
before}(t)$, and this together with $\mbox{\boldmath $r$}^{\rm
before}(t) = \mbox{\boldmath $r$}^{\rm after}(t)$ implies
\begin{equation}
\mbox{\boldmath $R$}^{\rm after} = \mbox{\boldmath $R$}^{\rm before} 
                            + 2\mbox{\boldmath $u$}^{\rm before}(t).
\end{equation}
This is the change in \mbox{\boldmath $R$} produced by a transit.  We
choose to leave the unit vectors $\hat{\mbox{\boldmath
$w$}}_{\lambda}$ in Eq.\ (\ref{wchoice}) unaffected by transits,
leaving only the effect on the phases $\alpha_{\lambda}$ to be
determined.  They must change in such a way as to reverse the sign of
$\mbox{\boldmath $u$}(t)$ but conserve $\mbox{\boldmath $v$}(t)$;
since $\mbox{\boldmath $u$}(t)$ is a sum of sines while
$\mbox{\boldmath $v$}(t)$ is a sum of cosines, this is easily done by
reversing the signs of the arguments $(\omega_{\lambda}t +
\alpha_{\lambda})$ in Eq.\ (\ref{rt}).  Let the transit occur at time
$t_{0}$; then $\omega_{\lambda}t_{0} + \alpha_{\lambda}^{\rm after} =
-(\omega_{\lambda}t_{0} + \alpha_{\lambda}^{\rm before})$ so
\begin{equation}
\alpha_{\lambda}^{\rm after} = -2\omega_{\lambda}t_{0} -
                         \alpha_{\lambda}^{\rm before}.
\end{equation}
Thus, a transit is implemented at time $t_{0}$ by leaving the
$\hat{\mbox{\boldmath $w$}}_{\lambda}$ alone and making the substitutions
\begin{eqnarray}
\mbox{\boldmath $R$} & \rightarrow & \mbox{\boldmath $R$} + 2\mbox{\boldmath
                                                 $u$}(t_{0}) \nonumber \\
\alpha_{\lambda} & \rightarrow & -2\omega_{\lambda}t_{0} - \alpha_{\lambda}.
\label{imptrans}
\end{eqnarray}
This conserves {\boldmath $r$}$(t)$, reverses the sign of {\boldmath
$u$}$(t)$, and conserves {\boldmath $v$}$(t)$.

Now the model consists of two parts.  (a) Between transits, the
average particle moves nondiffusively as given by Eq.\ (\ref{rt}) and
(\ref{wchoice}), with the phases $\alpha_{\lambda}$ and unit vectors
$\hat{\mbox{\boldmath $w$}}_{\lambda}$ assigned randomly.  (b) In
each small time interval $\Delta t$ a transit occurs with probability $\nu
\Delta t$; if it occurs, it replaces \mbox{\boldmath $R$} and the
$\alpha_{\lambda}$ with new values according to Eq.\ (\ref{imptrans}).
With the addition of transits, we can no longer express {\boldmath
$r$}$(t)$ and {\boldmath $v$}$(t)$ in closed form at all times, so we
no longer have a closed form for $Z(t)$; but the model can be
implemented easily on a computer, and then the data from the run can
be used to calculate $Z(t)$ in a manner analogous to an MD simulation.
We turn to comparison of the predictions of this model with MD results
next.

\section{Comparison with MD}
\label{MD}

The MD setup used to test our model is that described in \cite{wall4}
with two changes: $N = 500$ in all runs and the MD timestep was
reduced to $\delta t = 0.2 t^*$, where $t^* = 7.00 \times 10^{-15}$ s
is the natural timescale defined in \cite{wall4}.  (The system's mean
vibrational period $\tau = 2\pi/\omega_{\rm rms}$, where $\omega_{\rm
rms}$ is the rms average of the normal mode frequency distribution, is
approximately $287\,\delta t$.)  We performed equilibrium runs of the
system at 6.69 K, 22.3 K, 216.3 K, 309.7 K, 425.0K, 664.7 K, and
1022.0 K; at the lower two temperatures the system is not diffusing
(as can be seen from the system's mean square displacement, or from
the integral of $Z(t)$), and the system is diffusing otherwise.  Since
$T_{m} = 371.0$ K for Na at this density, our simulations range from
the glassy regime to nearly three times the melting temperature.  We
then ran the model for various values of $\nu$, adjusting until the
model matched the value of the first minimum of $Z(t)$ at each
temperature.  The values of $\nu$ that we fit for all temperatures are
given below; Figs.\ \ref{MDvsmodel} through \ref{1022.0Kvsmodel}
compare the model's predictions with the MD results for 
$\hat{Z}(t)= Z(t)/Z(0)$.  The model requires $\nu = 0$ for both 
nondiffusing states, so they are presented together in Fig.\ \ref{MDvsmodel}.
\begin{center}
\begin{tabular}{rl}
$T$ (K) &  $\nu \ (\tau^{-1})$ \\ \hline
 6.69   & 0.0     \\
 22.3   & 0.0     \\
 216.3  & 0.35018 \\
 309.7  & 0.60276 \\
 425.0  & 0.83985 \\
 664.7  & 1.24858 \\
1022.0  & 1.68774
\end{tabular}
\end{center}
\begin{figure}[p]
\includegraphics{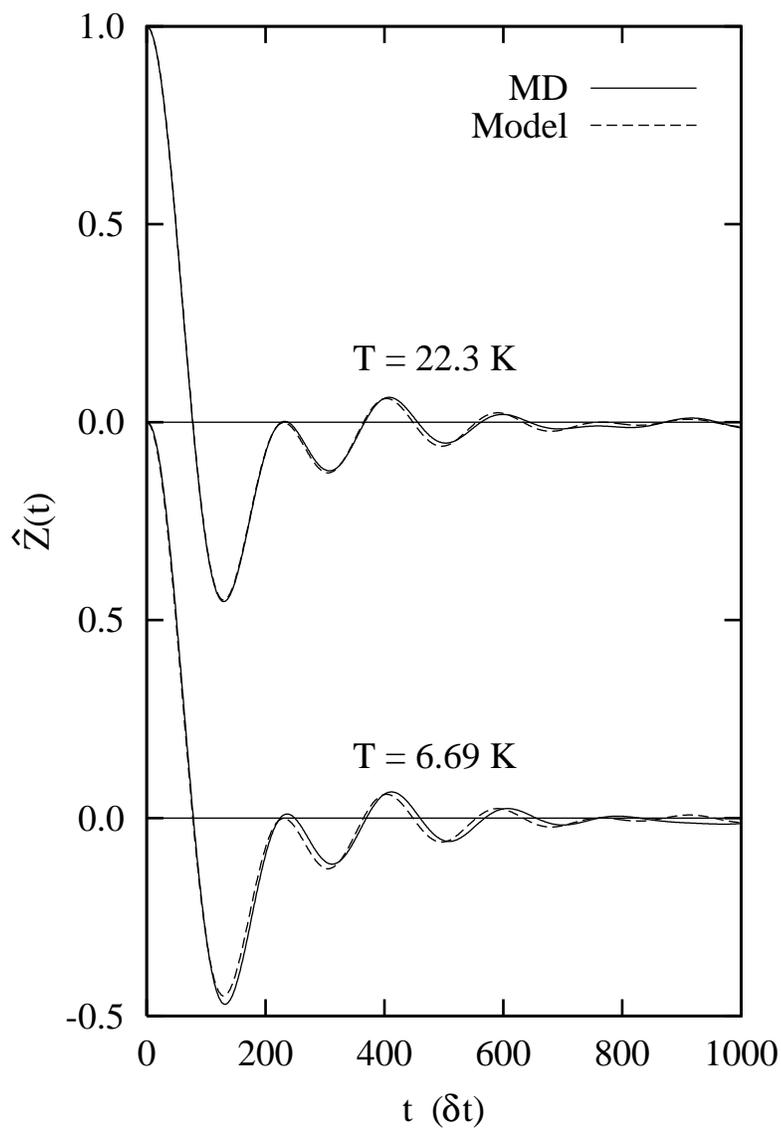}
\caption{The model prediction for $\hat{Z}(t)$ at $\nu = 0.0$ compared with 
         the MD results for glassy liquid Na at $T = 6.69$ K and $T = 22.3$K.}
\label{MDvsmodel}
\end{figure}            
\begin{figure}[p]
\includegraphics{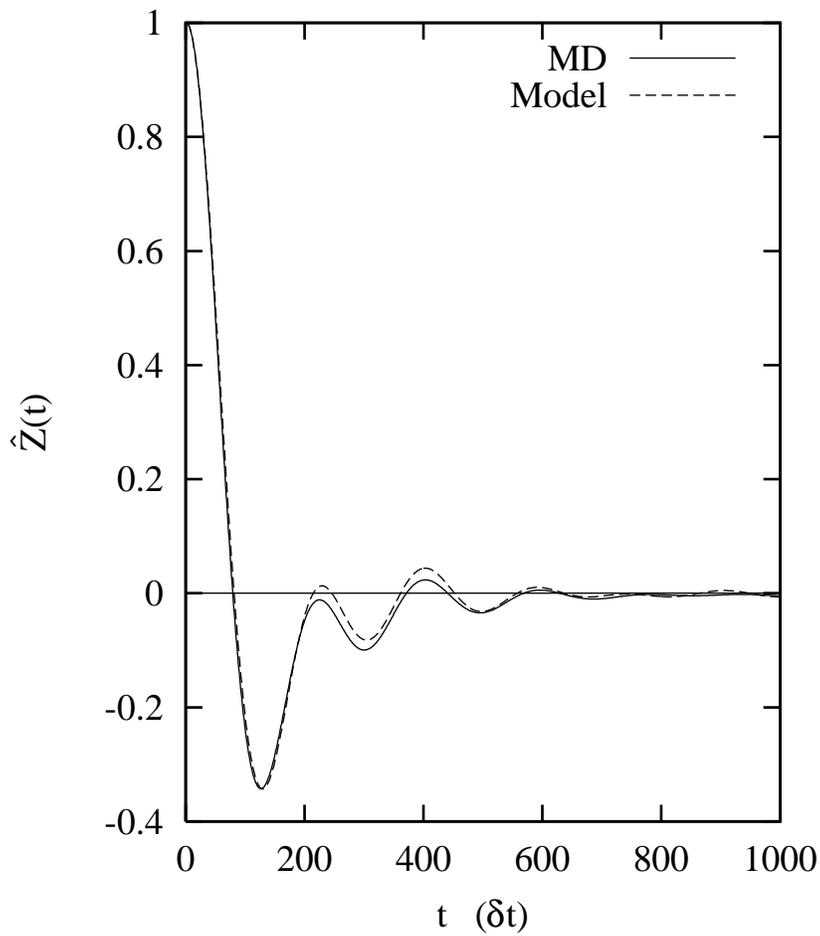}
\caption{The model prediction for $\hat{Z}(t)$ at $\nu = 0.35018 \, \tau^{-1}$
         compared with the MD result for supercooled liquid Na at $T =
         216.3$ K.}
\label{216.3Kvsmodel}
\end{figure}                       
\begin{figure}[p]
\includegraphics{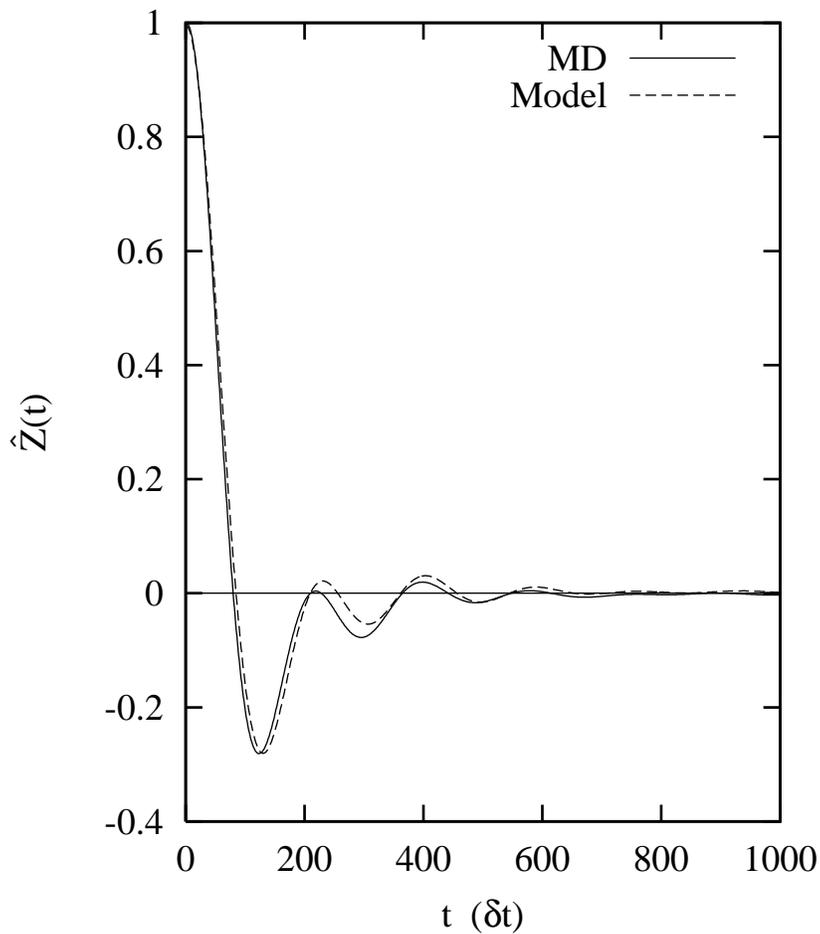}
\caption{The model prediction for $\hat{Z}(t)$ at $\nu = 0.60276 \, \tau^{-1}$
         compared with the MD result for supercooled liquid Na at $T =
         309.7$ K.}
\label{309.7Kvsmodel}
\end{figure}
\begin{figure}[p]
\includegraphics{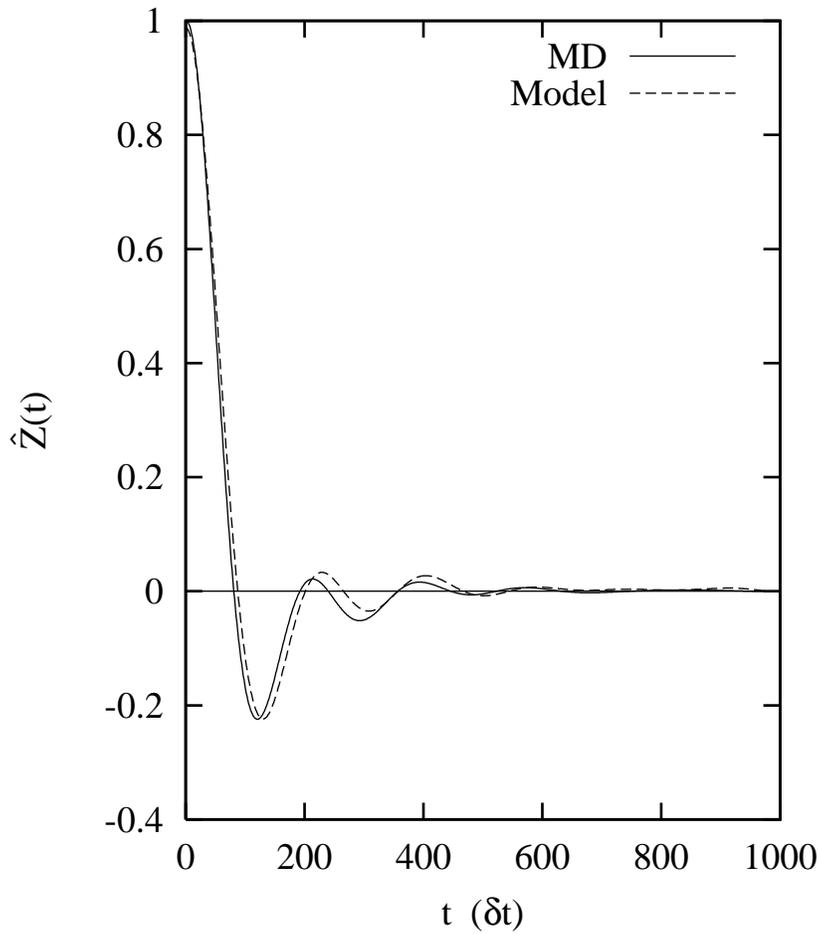}
\caption{The model prediction for $\hat{Z}(t)$ at $\nu = 0.83985 \, \tau^{-1}$
         compared with the MD result for liquid Na at $T = 425.0$ K.}
\label{425.0Kvsmodel}
\end{figure}
\begin{figure}[p]
\includegraphics{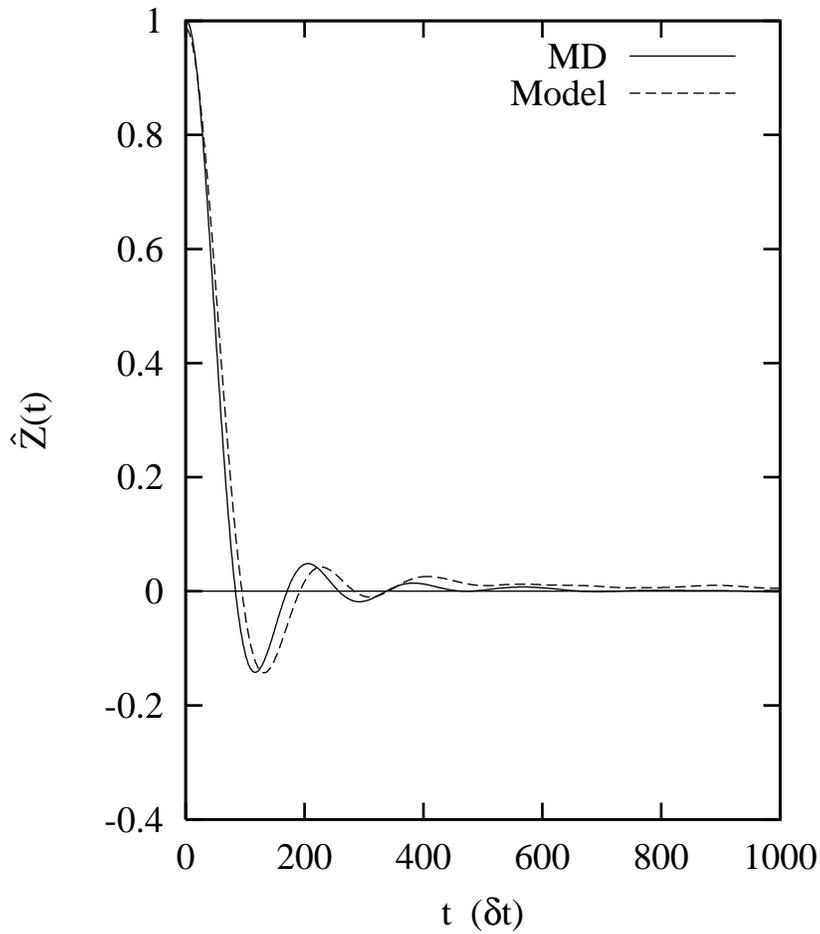}
\caption{The model prediction for $\hat{Z}(t)$ at $\nu = 1.24858 \, \tau^{-1}$
         compared with the MD result for liquid Na at $T = 664.7$ K.}
\label{664.7Kvsmodel}
\end{figure}
\begin{figure}[p]
\includegraphics{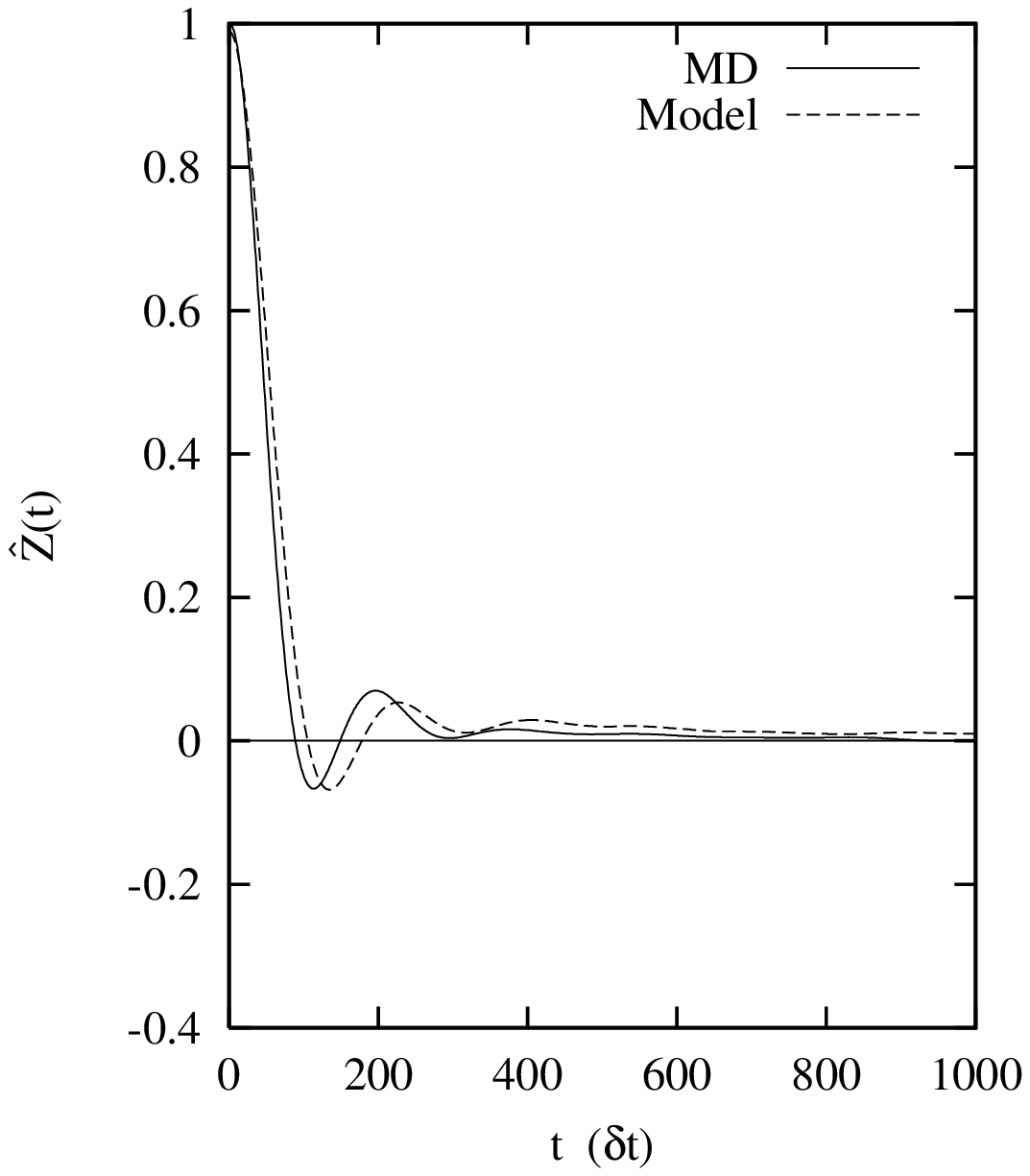}
\caption{The model prediction for $\hat{Z}(t)$ at $\nu = 1.68774 \, \tau^{-1}$
         compared with the MD result for liquid Na at $T = 1022.0$ K.}
\label{1022.0Kvsmodel}
\end{figure} 
Notice that in all diffusing cases $\nu$ is of the same order of
magnitude as $\tau^{-1}$, indicating roughly one transit per mean
vibrational period, as predicted in \cite{wall6} and noted in Section 
\ref{intro}.

\begin{figure}[p]
\includegraphics{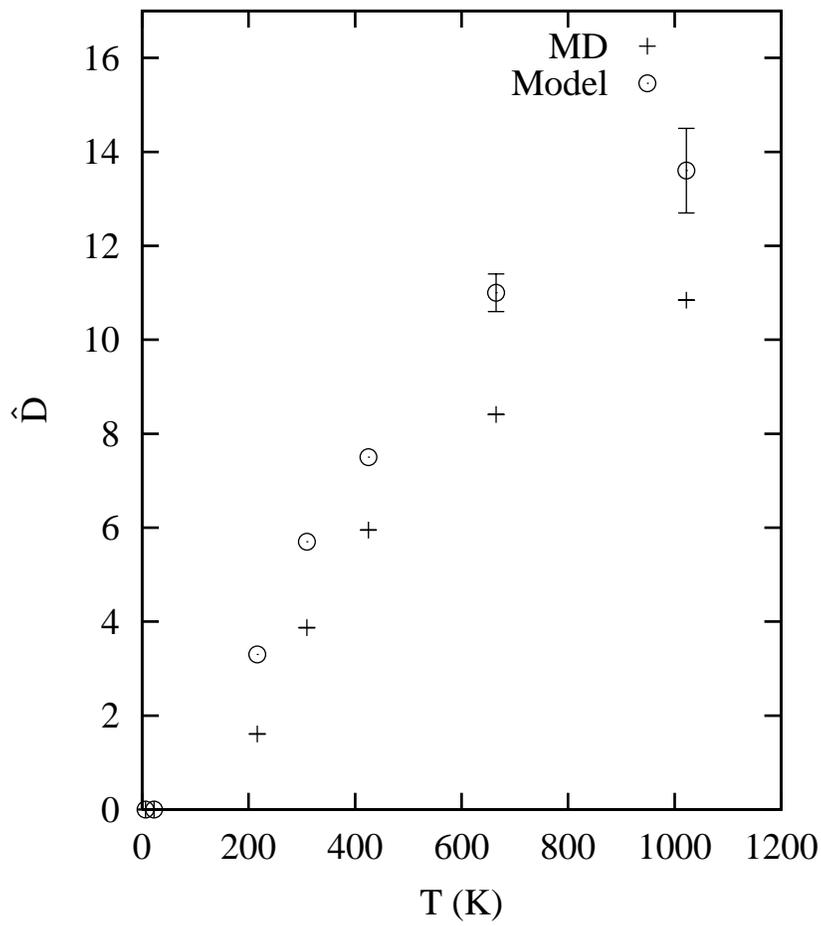}
\caption{$\hat{D}$ as a function of $T$ for both the model and MD.}
\label{DvsT}
\end{figure}

The most obvious trend in $\hat{Z}(t)$ is that its first minimum is
rising with increasing $T$; this is the primary reason for the
increasing diffusion coefficient $D$.  Note that the model is able to
reproduce this most important feature quite satisfactorily.  In fact,
all fits of the model to the MD results capture their essential
features, but we do see systematic trends in the discrepancies.
First, note that the location of the first minimum barely changes at
all in the model as $\nu$ is raised, but in MD the first minimum moves
steadily to earlier times as the temperature rises.  The first minimum
occurs at a time roughly equal to half of the mean vibrational period
(recall $\tau = 287 \, \delta t$), so the steady drift backward
suggests that the MD system is sampling a higher range of frequencies
at higher $T$.  Also, for the three lowest diffusing temperatures the model
tends to overshoot the MD result in the vicinity of the first two
maxima after the origin, and at the highest two temperatures this
overshoot is accompanied by a positive tail that is slightly higher
than the (still somewhat long) tail predicted by MD.  These overshoots
should clearly affect the diffusion coefficient $D$.  To check this, we 
calculated the reduced diffusion coefficient $\hat{D}$, the integral
of $\hat{Z}(t)$, which is related to $D$ by $D = (kT/M) \hat{D}$.  The
results are compared to the values of $\hat{D}$ calculated from the MD
runs in Fig.\ \ref{DvsT}.
In all of the diffusing cases, the model overestimates $\hat{D}$ by roughly
the same amount, which we take to be the effect of the overshoots at the first
two maxima.  At the higher temperatures the discrepancy is also higher,       
presumably due to the model's long tail. 

\section{Conclusions}
\label{concl}

We have presented a single-atom model of a monatomic liquid that
provides a unified account of diffusing and nondiffusing behavior.
The nondiffusing motion is modeled as a sum of oscillations at the
normal mode frequencies (Eq.\ (\ref{rt}) and (\ref{wchoice}));
self-diffusion is accounted for in terms of instantaneous transits
between wells, which occur at a temperature-dependent rate $\nu$.
\mbox{Since} this model gives a simple and straightforward account of the
motion itself, it can easily be used to calculate any single-atom
correlation function one wishes; here we have focussed on $Z(t)$ and
its integral $D$.  The relaxation of correlations expressed by the
decay of $Z(t)$ arises here from two distinct processes: Dephasing as
a result of the large number of frequencies in the single-valley motion,
and transits between valleys.  The dephasing effect produces relaxation
but not diffusion:  It causes $Z(t)$ to decay but its integral remains 
zero.  On the other hand, transits certainly contribute to relaxation, 
but in addition they raise the first minimum of $Z(t)$ substantially, 
increasing its integral and providing a nonzero $D$.

Most other workers in the field have studied Lennard-Jones or
molecular liquids; the only other work with liquid Na we have found is
Wu and Tsay's INM analysis \cite{wutsay1,wutsay2}, with which our
results are of comparable quality.  This is remarkable in light of the
fact that our model of the transit process is exceedingly simple; one
would expect that a more realistic model would produce even better
results.  We are also pleased to see that the model retains its
validity from the glassy regime to well beyond the liquid's melting
temperature.

In light of these results, answers to the following questions are
worth pursuing.  Do other monatomic liquids exhibit the same division
of their potential valleys into random and symmetric that liquid Na
does?  (It is known that LJ Ar does \cite{wall7}.)  How harmonic are
these valleys?  How similar are their frequency distributions?  Fig.\
\ref{MDvsmodel} shows that in Na the valleys are very nearly perfectly
harmonic, and we expect the same qualitative potential surface for all
nearly-free-electron metals (a total of 24 elemental liquid metals),
but other liquids might show more pronounced anharmonicities, and
those would need to be accounted for in the model.  Could a more
sophisticated model of the transit process (as opposed to simply
transiting forward) produce the shift in the first minimum and smaller
long-time tail shown in MD?  Can one develop a theory to predict the
transit probability $\nu$?  (Such a theory would be conceptually
related to the decay of the cage correlation function of
\cite{rabgezber2,rabgezber4}.)  Finally, how can these ideas be
applied to theories of other transport coefficients, such as bulk and
shear viscosities?  Future work will focus on answering these
questions.

\begin{center} \bf Acknowledgements \end{center}
This work was supported by the U.~S. DOE through contract W-7405-ENG-36.

\end{document}